\documentclass[aps,prb,onecolumn,groupedaddress,showpacs,showkeys]{revtex4}

\usepackage{amsmath}
\usepackage{amsfonts}
\usepackage{bm}
\usepackage{graphicx}
\usepackage{color}
\usepackage{subfigure}
\usepackage{hyperref}

\newcommand {\si}{\mbox{\boldmath$\sigma$}}

\begin{document}

\title{R-matrix formalism for electron scattering in two dimensions}

\author{P. N. Racec}
\email{racec@wias-berlin.de}
\affiliation{Weierstra\ss -Institut f\"ur Angewandte Analysis und Stochastik,
Mohrenstr. 39 10117 Berlin, Germany }
\affiliation{National Institute of Materials Physics, PO Box MG-7,
             077125 Bucharest Magurele, Romania}

\author{E.\ R.\ Racec}
\email{roxana@physik.tu-cottbus.de}
\affiliation{Institut f\"ur Physik, Technische Universit\"at Cottbus, Postfach 101344, 03013 Cottbus, Germany }
\affiliation{Faculty of Physics, University of Bucharest, PO Box MG-11,
             077125 Bucharest Magurele, Romania}

\author{H. Neidhardt}
\email{neidhard@wias-berlin.de}
\affiliation{Weierstra\ss -Institut f\"ur Angewandte Analysis und Stochastik,
Mohrenstr. 39 10117 Berlin, Germany }

\begin{abstract}
We investigate the scattering phenomena in two dimensions produced by a
general finite-range nonseparable potential. 
This situation can appear either 
in a Cartesian geometry
or 
in a heterostructure with cylindrical symmetry.
Increasing the dimensionality of the scattering problem 
new processes as the scattering between conducting channels and
the scattering from conducting to evanescent channels
are allowed.
For certain values of the energy called resonance energy
the transmission through the scattering region changes
dramatically in comparison with an one-dimensional problem.
If the potential has an attractive character 
even the evanescent channels can be seen 
as dips of the total transmission.
The multi-channel current scattering matrix is determined
using its representation in terms of the R-matrix.
The resonant transmission peaks 
are characterized quantitatively through 
the poles of the current scattering matrix.
Detailed maps of the localization probability density sustain
the physical interpretation of the resonances. 
Our formalism is applied to
a quantum dot in a two-dimensional electron gas and
to a conical quantum dot dot embedded inside a cylindrical nanowire.
\end{abstract}



\maketitle

\section{Introduction}

There is a permanent requirement of shrinking the semiconductor devices
in integrated circuits \cite{itrs}.  
As feature sizes shrink into the nanometer scale
regime, the device behavior becomes increasingly complicated since new physical
phenomena at short distances occur and limitations in material properties
are reached.  In order to keep the good characteristics for
transistors, new transistor architectures
were developed progressively in the last decade. 

Nowadays, there are developments of planar MOSFETs (metal oxide semiconductor
field effect transistors) \cite{intel} as well as of 
gate-all-around (GAA) MOSFETs \cite{Samuelson_06,suk_iedm06,cho_08}.
Both 
systems are also strongly related to more fundamental research
structures developed in the last years, like
in-plane-gate transistors \cite{wieck90}, 
single-electron transistors \cite{goeres},
silicon-on-insulator planar double-gate transistors \cite{wong_iedm97},
non-planar double-gate FinFETs \cite{huang_iedm99}, 
non-planar trigate transistors \cite{doyle03},
nanowire-based field-effect transistors (FET) \cite{Lieber_06},
nanowire resonant tunneling diodes \cite{samuelson02,wensorra},
nanowire lasers \cite{Lieber_08},
 or nanowire qubits \cite{Lieber_qubit},
whose maturity has still to be proven for industrial applications.
Their structural complexity has also progressively increased,
allowing for double-barrier structures \cite{samuelson02,wensorra}, 
or multiple core-shell layers \cite{Lieber_07,Lieber_08}.
The material composition includes 
mainly 
III-V materials 
GaAs/AlGasAs \cite{wieck90,goeres,wensorra}, 
InAs/InP \cite{samuelson02},
GaN/InGaN \cite{Lieber_08},
but also
group IV materials
Si \cite{suk_iedm06,suk_iedm06_ball}, 
and Si/Ge \cite{Lieber_06,Lieber_qubit},
predominant in the industry.

The transport phenomena in these mesoscopic devices
go beyond the semi-classical limit, and a
quantum mechanical description of the current and charge
densities \cite{vasileska_08} is necessary. The
most appropriate method for analyzing semiconductor devices with an active 
region in the nanometer scale and which are almost open (i.e. showing a
strong coupling between the active region and contacts)
is the scattering theory.

This paper is focused on 
systems for which the scattering process is a 
two-dimensional one. Such systems can appear either in a Cartesian geometry,
like for devices tailored in a two-dimensional electron gas (2DEG)
\cite{wieck90, goeres}, 
or in a cylindrical geometry, like for nanowire-based devices
\cite{Samuelson_06,suk_iedm06,cho_08,samuelson02,wensorra}.
In these systems there is a strong confinement of the motion 
in one direction, called \textit{transversal direction},
while the transport occurs in the other direction, called
\textit{longitudinal direction}.
The scattering problem is a two-dimensional one because the scattering 
potential is nonseparable, and also
the incoming electrons can choose different energy channels for 
transport which are mixed due to the scattering. 

We present in this work a general method,
valid within the effective mass approximation,
for solving the two-dimensional (2D) Schr\"odinger equation with scattering
boundary conditions.
Its solutions are found using the scattering theory and 
the R-matrix formalism
\cite{smr,wulf98,onac,roxana01,racec02,nemnes04,nemnes05,jayasekera06,
07impe,hagen08,milnikov09}.
This method is a semi-analytical one, and it gives the 
scattering functions in each point inside and outside the scattering area 
and for each energy as a function of the solutions of the Wigner-Eisenbud
problem. 
As known from the nuclear physics, the Wigner-Eisenbud problem
is the eigenvalue problem of the Hamilton operator 
for the closed counterpart of the considered open quantum system \cite{wigeis}.
The R-matrix formalism is not only numerically very efficient, but 
it is also suitable for higher dimensional nanostructures
with complex geometry \cite{wulf98,onac,roxana01,nemnes04,nemnes05}
and general nonseparable scattering potential \cite{racec09,roxana09}.
It can also deal with more than two terminals \cite{jayasekera06,milnikov09}.

Using the scattering functions we analyze further the transport properties
of the open quantum structures, especially the conductance. Besides the low 
dimensionality of such systems, the open character is also 
an essential feature which controls the transport phenomena
through the structure. When the quantum system becomes open,
its eigenstates yield resonance states which do not have an infinite life 
time anymore and which are not strictly localized inside the quantum system.
In this paper we identify the signature of the resonance in each 
conductance peak and study the influence of the nonseparable character 
of the potential on the resonances and on the conductance 
through the system.

An interesting effect in a multi-channel scattering problem is that
as soon as
the potential is not separable anymore, the channels get mixed.
If furthermore the scattering potential is attractive,
then it leads to unusual scattering properties, 
like resonant dips in the transmission
coefficient just below the next channel minimum energy. 
As it was shown analytically 
for a $\delta$ scattering potential \cite{bagwell90} and 
later on for a finite-range scattering potential \cite{levinson93,noeckel94} 
the dips are due to the quasi-bound-states splitting off from a higher 
evanescent channel. 
So that evanescent channels can not be neglected
when analyzing scattering in two- or three-dimensional quantum systems. 
These findings were recently confirmed numerically for a 
Gaussian-type scatterer \cite{gudmundsson04} and also for 
a quantum dot or a quantum ring \cite{gudmundsson05} embedded inside
nanowires tailored in a two-dimensional electron gas (2DEG),
or inside cylindrical nanowires \cite{racec09}.
The high resolution maps for the "near field" scattering
wave functions presented in Refs. \cite{gudmundsson04,gudmundsson05,
racec09}
show explicitly increased localization probability around the scatterer
for energies of the quasi-bound states, 
in agreement with the resonant reflection or resonant back-scattering
interpretation of these dips \cite{bagwell90,levinson93}.
The Cartesian and cylindrical geometries present 
different "selection rules" for the intersubband transmission \cite{racec09}.

\section{Model}
\label{model}

The electronic states in mesoscopic systems are easily described
within the effective mass approximation whose validity requires
that the envelope function $\Psi(E,\vec{r})$ must be slowly varying over
dimensions comparable to the unit cell of the crystal \cite{bastard}.

In the spherical effective mass approximation, 
the envelope function associated to the energy $E_{3D}$ 
satisfies a Schr\"odinger-type equation
\begin{equation}
\left[ -\frac{\hbar^2}{2 m^*} \Delta + V(\vec{r}) \right] \Psi(\vec{r})
   =E_{3D} \Psi(\vec{r}).
\label{Schr_3D}
\end{equation}
The so-called \textit{scattering potential} $V(\vec{r})$
contains the information about the confinement in the transversal direction, 
and inside the allowed area it is
a sum of the heterojunction
conduction band discontinuities, the electrostatic potential
due to the ionized donors and acceptors,
the self-consistent Hartree and exchange potentials due to free
carriers, and external potentials.
We use the symbol $m^*$ to denote the effective mass of the electrons, 
while $m$ denotes the magnetic quantum number.

For systems tailored in the 2DEG, the growth direction is chosen 
the $z$-direction,
while the plane of the 2DEG is $(x,y)$. The wave function in the $z$-direction,
$\xi(z)$, 
is taken as known 
(the simplest form is provided by Fang and Howard, \cite{fang_howard66})
so that the three-dimensional (3D) wave function can be written as
\begin{equation}
\Psi(E_{3D};\vec{r})=\xi(z) \psi(x,y).
\end{equation}
The total energy
\begin{equation}
E_{3D}=E_{2DEG}+E,
\end{equation}
where $E_{2DEG}$ is the energy of the 2DEG level, and $E$ is the 
energy associated with the motion in the plane of the 2DEG.
The 3D Schr\"odinger type equation reduces to a 
two-dimensional Schr\"odinger equation \cite{roxana09}
\begin{equation}
\left[ -\frac{\hbar^2}{2 m^*}
        \left( \frac{\partial^2}{\partial x^2}
              +\frac{\partial^2}{\partial y^2}\
        \right)
      +V(x,y)
\right] \; \psi(x,y)
= E \; \psi(x,y), 
\qquad x \in (-\infty, \infty), y \in [-d_y,d_y]. 
\label{2D_schr_cart}
\end{equation}
In the transversal direction the electron motion is limited at the interval 
$[-d_y,d_y]$ by a confining potential that we have considered as infinite.
The nonseparable potential $V(x,y)$ varies strongly with the position only 
inside a small domain ($|x| \le d_x$, $|y| \le d_y$)  
which is usually called
\textit{scattering region} and is quasi-constant outside this domain.

For cylindrical nanowires, 
the azimuthal symmetry suggests to use
cylindrical coordinates, with $z$ axis along the nanowire \cite{racec09}. 
As long as there are not split gates
on the surface of the nanowire,
the potential energy $V(\vec{r})$ 
is rotational invariant
\begin{equation}
V(\vec{r})=V(r, z)
\end{equation}
and nonseparable 
inside the scattering region. 
A scattering potential which does not explicitly depend on the 
azimuthal angle $\theta$ imposes the eigenfunctions of the 
orbital angular momentum operator $L_z$ as solutions of Eq. 
(\ref{Schr_3D})
\begin{equation}
\Psi_{m}(E_{3D};r,\theta,z)=\zeta_m(\theta) \psi_{m}(r,z),
\label{Psi_cyl_gen}
\end{equation}
where 
\begin{equation}
\zeta_m(\theta)=\frac{e^{im\theta}}{\sqrt{2\pi}},
\label{zeta}
\end{equation}
and
$m =0, \pm 1, \pm 2, ...$ is the magnetic quantum number.
This is an integer number due to 
the requirement that the function $e^{im\theta}$ should be single-valued.
The functions $\psi_{m}(r,z)$
are determined from the equation
\begin{equation}
\left[ -\frac{\hbar^2}{2 m^*}
             \left( \frac{\partial^2}{\partial r^2}
                   +\frac{1}{r}\frac{\partial}{\partial r}
                   -\frac{m^2}{r^2}
                   +\frac{\partial^2}{\partial z^2}
             \right)
       + V(r,z)
\right] \psi_{m}(r,z)
 =  E  \psi_{m}(r,z),
\qquad r\in [0,R], z \in (-\infty, \infty), 
\label{2D_schr_cyl}
\end{equation}
where $E$ denotes here the kinetic energy associated with the 3D
motion of the electron inside the nanowire, $E=E_{3D}$.
We have also considered an infinite potential outside the nanowire.
In such a way, every magnetic quantum number $m$ defines a 
\textit{two-dimensional (2D) scattering problem}. 
Furthermore, these
2D scattering problems can be solved separately
if the scattering potential is rotational invariant. 
How many of these problems have to be solved, depends on the 
specific physical quantity which has to be computed.

\subsection{Scattering problem for two dimensions}

We consider the following Schr\"odinger type equation in two dimensions,
$(x_\parallel, x_\perp)$, denoting generically the longitudinal and
the transversal direction, respectively,
\begin{equation}
\left[ -\frac{\hbar^2}{2 m^*} \Delta_{x_\parallel, x_\perp}
       + V(x_\parallel, x_\perp)
\right] \psi(E;x_\parallel, x_\perp)
 =  E  \psi(E;x_\parallel, x_\perp), 
\qquad x_\perp \in \Omega, x_\parallel \in (-\infty, \infty). 
\label{2D_schr}
\end{equation}
One could consider here different effective masses in the longitudinal 
and transversal directions and also, for a layered heterostructure,
a position-dependent effective mass. These effects can be incorporated in the
formalism, but within this paper we neglect them for the simplicity of the 
exposure.

For the Cartesian geometry \cite{roxana09}, in comparison 
with Eq. (\ref{2D_schr_cart}) we have
\begin{gather}
x_\parallel= x, x_\perp=y , \Omega=[-d_y,d_y], \nonumber \\
\Delta_{x_\parallel, x_\perp} = \frac{\partial^2}{\partial x^2}
              +\frac{\partial^2}{\partial y^2}, \\
E=E_{3D}-E_{2DEG}, \nonumber
\end{gather}
while for the cylindrical geometry \cite{racec09},
in comparison with Eq. (\ref{2D_schr_cyl}) we have
\begin{gather}
x_\parallel= z, x_\perp=r , \Omega = [0,R], \nonumber \\
\Delta_{x_\parallel, x_\perp} = \frac{\partial^2}{\partial r^2}
                   +\frac{1}{r}\frac{\partial}{\partial r}
                   -\frac{m^2}{r^2}
                   +\frac{\partial^2}{\partial z^2}, \\
E=E_{3D} \nonumber,
\end{gather}
which depends on the magnetic quantum number $m$. 

The analogy between the both geometries appears more evident considering
for the cylindrical geometry the unitary transformation 
$U:L^2([0,R]\times \mathbb{R}, rdrdz) \rightarrow 
   L^2([0,R]\times \mathbb{R}, drdz)$,
with $Uf(r,z)=g(r,z)=\sqrt{r}f(r,z)$. 
The inverse transformation is
$U^\dagger:L^2([0,R]\times \mathbb{R}, drdz) \rightarrow 
           L^2([0,R]\times \mathbb{R}, rdrdz)$,
with $U^\dagger g(r,z)=f(r,z)=(1/\sqrt{r})g(r,z)$.
In such a way, the Schr\"odinger operator becomes
\begin{eqnarray}
\tilde{H}_{r,z} & = & U H_{r,z}U^\dagger=
U \left[ -\frac{\hbar^2}{2 m^*} \Delta_{z,r} +V(z,r) \right] U^\dagger 
\nonumber \\
& = & -\frac{\hbar^2}{2 m^*} \left[ \frac{\partial^2}{\partial r^2}
                   -\frac{m^2-1/4}{r^2}
                   +\frac{\partial^2}{\partial z^2}\right] +V(z,r).
\end{eqnarray}
In turn, the term which contains $1/r^2$ and which is specific 
for the cylindrical geometry
plays the role of a potential, and the Laplace operator appears
as known for two dimensions in Cartesian coordinates. 
The current scattering matrix and the localization probability distribution
density 
of an electron in a scattering state remain unchanged under this unitary
transformation.

\begin{figure}[h]
\begin{center}
\includegraphics*[width=4.in]{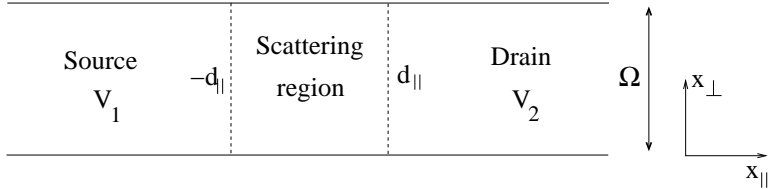}
\caption{The generic geometry of the 2D scattering problem.
}
\label{Schr_2D_sketch}
\end{center}
\end{figure}

Due to the localized character of the scattering potential
it is appropriate to solve Eq. (\ref{2D_schr}) within 
the scattering theory. 
The potential energy which appears in Eq. (\ref{2D_schr})
has generally two components:
\begin{equation}
V(x_\parallel, x_\perp) = V_\perp(x_\perp) + V_{scatt}(x_\parallel, x_\perp).
\end{equation}
The first one, $V_\perp(x_\perp)$, describes the lateral confinement 
of the electrons
and is translation invariant along the parallel direction $x_\parallel$.
We consider a hard wall potential
\begin{equation}
V_\perp(x_\perp) = \begin{cases}
              0, &      x_\perp \in \Omega \\
              \infty, & x_\perp \not\in  \Omega
            \end{cases},
\label{conf_hw}
\end{equation}
which defines a quantum wire for the Cartesian geometry and a cylindrical
nanowire for the second considered geometry.
A parabolic wall like in Ref. \cite{gudmundsson04,gudmundsson05}
may also be considered.

The scattering potential energy inside the nanowire, 
$V_{scatt}(x_\parallel, x_\perp)$,
has generally a nonseparable character in a domain of finite-range
and is constant outside this domain. 
We consider here the nonseparable potential 
localized within the area $\Omega \times [-d_\parallel,d_\parallel]$,
see Fig. \ref{Schr_2D_sketch},
\begin{equation}
V_{scatt}(x_\parallel, x_\perp)=\begin{cases}
           V_1, & x_\perp \in \Omega, x_\parallel < -d_\parallel \\
           W(x_\parallel, x_\perp), & x_\perp \in \Omega, 
                               -d_\parallel \le x_\parallel \le d_\parallel \\
           V_2, & x_\perp \in \Omega, x_\parallel > d_\parallel
         \end{cases}.
\end{equation}
There are not material definitions for the interfaces 
$x_\parallel = \pm d_\parallel$.
Usually, they are chosen inside the highly doped regions
of the heterostructure characterized by a slowly variation of the potential
in the longitudinal direction,
practically by a constant potential. These regions play the role 
of the source and drain contacts.

\subsection{Scattering states}

In the asymptotic regions,
 $|x_\parallel|> d_\parallel$ i.e. source and drain contacts,
the potential energy
is separable in the transversal (i.e. confinement) and 
the longitudinal (i.e. transport) direction,
i.e. $V(x_\parallel, x_\perp)=V_\perp(x_\perp)+V_s$, $s=1,2$,
and Eq. (\ref{2D_schr}) can be directly solved 
using the separation of variables method
\begin{equation}
\psi(E;x_\parallel, x_\perp)=\phi(x_\perp)\varphi(x_\parallel).
\label{sepxiphi}
\end{equation}
The function $\phi(x_\perp)$ satisfies the transversal equation
\begin{equation}
\left[
-\frac{\hbar^2}{2 m^*} \Delta_{x_\perp} +V_\perp(x_\perp)
\right]
\phi(x_\perp)=E_\perp \phi(x_\perp), \qquad x_\perp \in \Omega
\label{radeq}
\end{equation}
where
\begin{equation}
\Delta_{x_\perp}=\left\{ \begin{array}{ll}
                 \frac{d^2}{dy^2}, & \quad  \mbox{Cartesian geometry} \\
                 \frac{d^2}{dr^2}
                  + \frac{1}{r} \frac{d}{dr}
                  - \frac{m^2}{r^2}, & \quad \mbox{cylindrical geometry}
                 \end{array} \right. .
\end{equation}
The hard wall confinement potential requires
Dirichlet boundary condition at the boundaries $\partial \Omega$ 
of the interval $\Omega$, $\phi(\partial \Omega)=0$. 
As a remark, for the cylindrical
geometry, the boundary $r=0$ is an artificial one introduced in order to
use the cylindrical symmetry. At this boundary it is sufficient that 
$\phi(x_\perp)$ remains finite.

Due to the electron confinement in the transversal direction $x_\perp$
the solutions of Eq. (\ref{radeq})
define the \textit{transversal modes}, $\phi_n(x_\perp)$,
with the corresponding transversal energies $E_{\perp n}$,
$n\ge 1$.
The eigenfunctions $\phi_n(x_\perp)$ 
depend on the geometry (Cartesian or cylindrical) 
and on the confinement potential.
In the case of a hard wall confinement, the transversal modes 
are given for 
the Cartesian geometry by sine functions \cite{roxana09},
while for the cylindrical geometry 
they are expressed in terms of the Bessel functions of the first 
kind \cite{racec09}.
The transversal modes form an orthonormal and complete system of functions.

The function $\varphi(x_\parallel)$ satisfies the one-dimensional 
Schr\"odinger type equation called longitudinal equation
\begin{equation}
\left[ -\frac{\hbar^2}{2 m^*} \frac{d^2}{dx_\parallel^2} 
+ V_s \right] \varphi(x_\parallel)=
(E-E_\perp) \varphi(x_\parallel), 
\quad x_\parallel \in (-\infty, -d_\parallel) \cup (d_\parallel, \infty),
\label{transpeq}
\end{equation}
where $s=1$ stays for the source contact ($ x_\parallel < -d_\parallel $)
and $s=2$ for the drain contact ($x_\parallel > d_\parallel$).
In the case of different effective masses in transversal and 
longitudinal direction, one can use the corresponding effective mass
in each of the above equations.

Every transversal mode 
together with the associated motion on the transport direction
defines a \textit{scattering channel} 
on each side of the scattering area. 
The scattering channels are indexed by $(sn)$, $n\ge 1$, $s=1,2$ for each $E$.
In contradistinction to the Cartesian geometry, 
in the case of a cylindrical geometry 
there is a set of 
2D scattering problems,
indexed by the magnetic quantum number $m$, that 
have to be solved.
Consequently the scattering channels  
should be also indexed by $m$. For simplicity we omit
the index $m$ in this section, but we keep in mind that we solve here a 
2D scattering problem and obtain the scattering functions for a 
fix value of $m$.

If the total energy $E$ and the lateral eigenenergy $E_{\perp n}$
are fixed, 
there are at most two linearly independent solutions of Eq.
(\ref{transpeq}). In the asymptotic region
they are given as a linear combination of
exponential functions
\begin{equation}
\varphi_{sn}(x_\parallel) = \begin{cases}
                           A_s e^{ i k_{1n} x_\parallel}
                          +B_s e^{-i k_{1n} x_\parallel} , 
                                        & x_\parallel < -d_\parallel \\
                           C_s e^{ i k_{2n} x_\parallel}
                          +D_s e^{-i k_{2n} x_\parallel} , 
                                        & x_\parallel  > d_\parallel
                        \end{cases}
\label{psi_plane}
\end{equation}
where $A_s$, $B_s$, $C_s$ and $D_s$ are complex coefficients 
depending on $n$ and $E$ for each value of $s=1,2$.
The wave vector is defined for each scattering channel $(sn)$ as
\begin{equation}
k_{sn}(E) = k_0 \sqrt{(E-E_{\perp n}-V_s)/u_0},
\label{ksn}
\end{equation}
where $k_0=\pi/2d_\parallel$ and $u_0 = \hbar^2 k_0^2/2 m^*$.
In the case of the conducting or open channels 
\begin{equation}
E-E_{\perp n}-V_s \ge 0,
\label{open_channels}
\end{equation}
$k_{sn}$ are positive real numbers
and correspond to propagating plane-waves.
For the evanescent or closed channels 
\begin{equation}
E-E_{\perp n}-V_s < 0,
\label{closed_channels}
\end{equation}
$k_{sn}$  are
given from the first branch of the complex square root function,
$k_{sn}=i |k_{sn}|$,
and describe exponentially decaying functions away from the scattering
region.
Thus, the number of the conducting channels, $N_{s}(E)$, $s=1,2$,
is a function of energy, and for a fixed energy $E$
this is the largest value of $n$, which satisfies the
inequality (\ref{open_channels})
for given values of $s$. 

Each conducting channel corresponds to one degree of freedom for the electron
motion through the system and, consequently, there exists only one 
independent solution of Eq. (\ref{2D_schr}) for a fixed channel $(sn)$
associated with the energy $E$, $\psi^{(s)}_{n}(E;x_\parallel, x_\perp)$.
For describing further the transport phenomena in the frame of
the scattering theory it is convenient to consider this solution
as a \textit{scattering state}, i.e. as a sum of an incoming
component on the channel $(sn)$ and a linear combination of outgoing
components on each scattering channel. 
One can write the scattering wave functions in a compact form \cite{roxana09}
\begin{equation}
\psi^{(s)}_n (E;x_\parallel, x_\perp)
  =  \frac{\theta(N_{s}(E)-n)}{\sqrt{2\pi}}
  \begin{cases}
     \delta_{s1} e^{i k_{1n}(x_\parallel + d_\parallel)} \phi_n(x_\perp) +
             \sum_{n'=1}^\infty\limits S_{1n',sn}(E) 
                                  e^{-i k_{1n'}(x_\parallel+d_\parallel)} 
                           \phi_{n'}(x_\perp),
   & x_\parallel < - d_\parallel \\
     \delta_{s2} e^{-i k_{2n'}(x_\parallel-d_\parallel)} \phi_n(x_\perp) +
             \sum_{n'=1}^\infty\limits S_{2n',sn}(E) 
                                       e^{i k_{2n'}(x_\parallel-d_\parallel)} 
                           \phi_{n'}(x_\perp),
  & x_\parallel > d_\parallel
 \end{cases}
\label{scatt_funct}
\end{equation}
The step function $\theta$ 
in the above expressions, with 
$\theta(x\ge 0)=1$ and $\theta(x< 0)=0$,
assures that the
scattering functions are defined only for the conducting channels. 
Writing explicitly the position of the interfaces $\pm d_\parallel$
at the exponent has advantages for the analytical treatment of the scattering
problem \cite{roxana01,BKNR04b}.
As it is discussed in Refs. \cite{jayasekera06,racec09}, it is
necessary to consider the sum until infinity  in the second
term of the above expression, in order to keep the 
mathematical completeness of the transversal channels. 

The physical interpretation of the expressions (\ref{scatt_funct}) is
that,
due to the nonseparable character of the scattering potential,
a plane-wave incident onto the scattering domain is reflected on every
channel - open or closed for transport - on the same side of the system
and transmitted on every channel - open or closed for transport -
on the other side. The reflection and transmission amplitudes are described
by the complex coefficients $S_{sn',sn}$ and $S_{s'n',sn}$ with 
$s \ne s'$, respectively, and all of them should be nonzero. 
These coefficients define a matrix with $N_{1}(E)+N_{2}(E)$ 
infinite columns. For an elegant solution of the scattering problem we
extend $S(E)$ to an infinite square matrix and set at zero
the matrix elements without physical meaning,
$S_{s'n',sn}(E)=0$, $n > N_{s}(E)$, $s=1,2$.
In this way we define the \textit{wave transmission matrix}
or \textit{wave-function amplitudes matrix} \cite{bagwell90}. 
It is also called \textit{generalized scattering matrix} \cite{schanz95}.
This is not the well-known scattering matrix (current transmission
matrix) whose unitarity reflects the current conservation. 
The generalized scattering matrix is a non-unitary matrix,
which has the advantage that it allows for a description of the
scattering processes not only in the asymptotic region but also inside the
scattering area.

The three-dimensional scattering states, solutions of Eq. (\ref{Schr_3D}) 
can be now written as
\begin{equation}
\Psi^{(s)}_{n}(E;x_\parallel, x_\perp, x_3)=\omega(x_3)
\psi^{(s)}_{n}(E;x_\parallel, x_\perp),
\label{Psi_gen2}
\end{equation}
where $\omega(x_3)$ stays for $\xi(z)$ in the case of the Cartesian geometry and
for $\zeta_m(\theta)$ in the case of the cylindrical geometry. 
Being eigenfunctions of an open system,
the scattering states are ortho-normalized in the general sense \cite{07impe}
\begin{equation}
\int_\Omega d\Omega \int_{-\infty}^\infty dx_\parallel \;
\psi_{n}^{(s)}(E;x_\parallel, x_\perp) 
\psi_{n'}^{(s')}(E';x_\parallel, x_\perp)^*=
\delta_{ss'} \delta_{nn'} \frac{\delta(E-E')}{g_{sn}(E)},
\end{equation}
where 
$g_{sn}(E)=m^*/ [\hbar^2 k_{sn}(E)]$ is the 1D density of states. 
We have to mention that for the Cartesian coordinates \cite{roxana09}, 
the measures are $d\Omega=dy$, $dx_\parallel=dx$, while for the cylindrical 
geometry \cite{racec09} they are $d\Omega=r dr$, $dx_\parallel=dz$.

\subsection{R-matrix formalism for two dimensions}

The scattering functions inside the scattering region are determined
using the R-matrix formalism, i.e. they are 
expressed in terms of the eigenfunctions 
corresponding to the closed counterpart of the scattering
problem 
\cite{smr,wulf98,onac,roxana01,racec02,nemnes04,nemnes05,jayasekera06,
milnikov09}.
In our opinion this is a more appropriate method than the common mode 
space approach which implies the expansion of the scattering functions 
inside the scattering area in the basis of the transversal modes
$\phi_n(x_\perp)$. 
As it is shown in Ref. \cite{noeckel94},\cite{luisier_06}
the mode space approach has limitations
for structures with abrupt changes in the potential or
sudden spatial variations in the widths of the wire;
it breaks even down for coupling operators that are not
scalar potentials, like in the case of an external magnetic field.
In the R-matrix formalism the used basis contains all the information 
about the scattering potential, and this type of difficulties 
can not appear. 

Thus, the scattering functions inside the scattering region are given as
\begin{eqnarray}
\psi_{n}^{(s)}(E;x_\parallel, x_\perp)&=&\sum_{l=1}^\infty a_{ln}^{(s)}(E) \chi_{l}(x_\parallel, x_\perp),
\label{varphi_expan} 
\end{eqnarray}
with $x_\perp \in \Omega$ and 
$x_\parallel \in [-d_\parallel,d_\parallel]$. 

The so-called \textit{Wigner-Eisenbud functions}, 
$\chi_{l}(x_\parallel, x_\perp)$,
firstly used in the nuclear physics \cite{wigeis,lane},
satisfy the
same equation as $\psi_{n}^{(s)}(x_\parallel, x_\perp)$,
Eq. (\ref{2D_schr}), but with different  boundary conditions
in the transport direction. 
Since the scattering function
$\psi_{n}^{(s)}(x_\parallel, x_\perp)$ satisfies energy
dependent boundary conditions derived from Eq. (\ref{scatt_funct})
due to the continuity of the scattering function and its derivative at
$x_\parallel = \pm d_\parallel$, the Wigner-Eisenbud function
$\chi_{l}(x_\parallel, x_\perp)$ has to satisfy Neumann boundary
conditions at the interfaces between the scattering region and leads
\begin{equation}
\left. \frac{\partial \chi_{l}}{\partial x_\parallel}
\right|_{x_\parallel=\pm d_\parallel}=0, \quad l \ge 1.
\label{neumann_bc}
\end{equation}
The hard wall confinement potential requires
Dirichlet boundary condition at $\partial \Omega$
also for the Wigner-Eisenbud functions, 
$\chi_{l}(\partial \Omega,x_\parallel)=0$. 
As already mentioned for the scattering states, 
for the cylindrical geometry it is sufficient that the
Wigner-Eisenbud function remains finite at $r=0$. 
The functions $\chi_{l}$, $l \ge 1$, build a basis which verifies the
orthogonality relation
\begin{equation}
\int_\Omega d\Omega \int_{-d_\parallel}^{d_\parallel} dx_\parallel \;
                          \chi_l(x_\perp,x_\parallel) 
                          \chi_{l'}(x_\perp,x_\parallel)  
                            = \delta_{ll'}
\end{equation}
and the closure relation
\begin{equation}
\sum_{l=1}^\infty \chi_l(x_\perp,x_\parallel) 
                  \chi_l(x'_\perp,x'_\parallel)=\delta(x_\perp-x'_\perp)
                                           \delta(x_\parallel-x'_\parallel).
\label{closure_rel}
\end{equation}
Note that for the cylindrical geometry $\delta(x_\perp-x'_\perp)$
in the relation (\ref{closure_rel}) means
\mbox{$\delta(r-r')/r$}.
The corresponding eigenenergies to $\chi_{l}$ are denoted by $E_l$
and are called \textit{Wigner-Eisenbud energies}.
Since the Wigner-Eisenbud problem is defined on a closed volume
with self-adjoint boundary conditions, the 
eigenfunctions $\chi_{l}$ and the eigenenergies $E_l$ can be chosen 
as real quantities.
The Wigner-Eisenbud problem is, thus, the closed counterpart of the scattering
problem. 

In the case of the one-dimensional system
without spherical symmetry,
it was recently proven mathematically rigorous that 
the R-matrix formalism allows for a proper expansion of the scattering matrix
on the real energy axis \cite{hagen08}.
In this section we present an extension of 
the R-matrix formalism for 2D scattering problem.

To calculate the expansion coefficients $a_{ln}^{(s)}(E)$
we multiply Eq. (\ref{2D_schr}) by $\chi_{l}(x_\parallel, x_\perp)$ 
and the equation satisfied
by the Wigner-Eisenbud functions by $\psi_{n}^{(s)}(E;x_\parallel, x_\perp)$. 
The difference
between the resulting equations is integrated over
$\Omega \times [-d_\parallel,d_\parallel]$, 
with the corresponding measures, and one obtains on the right-hand side 
the coefficient $a_{ln}^{(s)}(E)$. 
After using the Green's theorem and the boundary conditions
one finds $a_{ln}^{(s)}(E)$ and feeds in it into
Eq. (\ref{varphi_expan}).
So,
\textit{the scattering functions inside the scattering region}
($x_\parallel \in [-d_\parallel,d_\parallel],\; x_\perp \in \Omega$)
are obtained in terms of their derivatives at the edges of this
domain,
\begin{equation}
\begin{split}
\psi_{n}^{(s)}(E;x_\parallel, x_\perp)
=\frac{1}{k_0} \int_\Omega d x'_\perp
 & 
   \left[ R(E;-d_\parallel,x'_\perp,x_\parallel,x_\perp)
   \left.
   \frac{\partial \psi_{n}^{(s)}(E;x'_\parallel,x'_\perp)}
        {\partial x'_\parallel} \right|_{x'_\parallel=-d_\parallel}
   \right. 
 \\
 &
  -\left. R(E;d_\parallel,x'_\perp,x_\parallel, x_\perp)
   \left.
   \frac{\partial \psi_{n}^{(s)}(E;x'_\parallel,x'_\perp)}
        {\partial x'_\parallel} \right|_{x'_\parallel=d_\parallel}
   \right], 
\end{split}
\label{psi-inside}
\end{equation}
where the $R$-function is defined as
\begin{equation}
R(E;x_\parallel, x_\perp,x'_\parallel, x'_\perp,) 
\equiv \frac{u_0}{k_0}\sum_{l=1}^\infty
  \frac{\chi_{l}(x_\parallel, x_\perp) \chi_{l}(x'_\parallel, x'_\perp)}
       {E-E_l}.
\label{R-function}
\end{equation}
The functions 
$\partial \psi_{n}^{(s)}/\partial x_\parallel$ at $x_\parallel=\pm d_\parallel$
are calculated from the asymptotic form (\ref{scatt_funct})
based on the continuity
conditions  for the derivatives of the scattering functions
on the interfaces
between the scattering region and leads.

With these results the scattering functions
inside the scattering domain
are expressed in terms of the
wave transmission matrix $\mathbf{S}$
\begin{equation}
\vec{\Psi}(E;x_\parallel, x_\perp) = \frac{i}{\sqrt{2 \pi}}  
                       \bm{\varTheta}(E) [{\bf 1} - \mathbf{S}^T(E)] \\
                   \mathbf{K}(E) \vec{R}(E;x_\parallel, x_\perp),
\label{psi3}
\end{equation}
where the component $(sn)$ of the vector $\vec{\Psi}$ is the
scattering function $\psi_{n}^{(s)}(E;x_\parallel, x_\perp)$, $n \ge 1$, $s=1,2$
and $\mathbf{S}^T$ denotes the matrix transpose.
The diagonal matrix $\mathbf{K}$ has on its diagonal the wave
vectors (\ref{ksn}) of each scattering channel 
\begin{equation}
\mathbf{K}_{sn,s'n'}(E)= \frac{k_{sn}(E)}{k_0} \, 
                               \delta_{nn'} \delta_{ss'},
\label{K-matrix}
\end{equation}
$n,n' \ge 1$, $s,s'=1,2$,
and the vector $\vec{R}(E;x_\parallel, x_\perp)$ is given as
\begin{equation}
\vec{R}(E;x_\parallel, x_\perp)= \frac{u_0}{\sqrt{k_0}}
                     \sum_{l=1}^{\infty}
                       \frac{\chi_{l}(x_\parallel, x_\perp) \vec{\chi}_{l}}
                            {E-E_l},
\label{R-vector}
\end{equation}
where $\vec{\chi}_{l}$ has the components
\begin{equation}
(\vec{\chi}_{l})_{sn} = \frac{1}{\sqrt{k_0}}
                       \int_\Omega \; 
                       \chi_{l}(x_\perp,(-1)^{s} d_\parallel) 
                       \phi_n(x_\perp) d\Omega ,
\label{chi-vector}
\end{equation}
$n \ge 1, s=1,2$. 
The diagonal $\varTheta$-matrix,
$\bm{\varTheta}_{sn,s'n'}(E) = \theta(N_{s}(E)-n) \,
                                 \delta_{ss'} \,
                                 \delta_{nn'}$, $n \ge 1$, $s=1,2$,
assures non-zero values only for the
scattering functions corresponding to the conducting channels.

Using further the continuity of the scattering functions on the surface of
the scattering area and expanding $\vec{R}(E;\pm d_\parallel,x_\perp)$ 
in the basis
$\left\{\phi_n(x_\perp)\right\}_{n \ge 1}$ we find the relation between
the matrixes $\mathbf{S}$ and $\mathbf{R}$
\begin{equation}
\mathbf{S}(E) = \left[ \mathbf{1} - 2 \left( \mathbf{1} + i \mathbf{R}(E) 
                                                        \mathbf{K}(E)
                                  \right)^{-1}
                \right] \bm{\varTheta}(E),
\label{R-Srelation}
\end{equation}
with the $R$-matrix given by means of a dyadic product
\begin{equation}
\mathbf{R}(E)= u_0 \sum_{l=1}^{\infty}
                     \frac{\vec{\chi}_{l} \, \vec{\chi}_{l}^T}
                          {E-E_l}.
\label{R-matrix}
\end{equation}
According to the above relation, $\mathbf{R}$ is an infinite-dimensional 
symmetrical real matrix and its elements 
defined by Eq. (\ref{R-matrix}) are dimensionless.
The above form allows for a very efficient
numerical implementation for computing the $R$-matrix. 

The expression (\ref{R-Srelation}) of the $S$-matrix in terms of the
$R$-matrix is the key relation for solving
2D scattering problems using only the eigenfunctions
and the eigenenergies of the closed quantum system. 
They contain the full information about the scattering potential
and carry it over to the $R$-matrix.
The matrix $\mathbf{K}$ characterizes
the contacts and can be constructed using only the information about
the potential in these regions.
On the base of  Eq. (\ref{R-Srelation}) the
wave transmission matrix
is calculated and after that the scattering functions in each point of the
system are obtained using
Eqs. (\ref{scatt_funct}) and (\ref{psi3}).
Further on, each transport property of the open quantum system can be derived 
from the scattering function in terms of the scattering matrix.

\subsection{Reflection and transmission coefficients}

Using the density current operator 
\begin{equation}
\vec{j}(\vec{r})=\frac{\hbar}{2i m^*}
  \Bigl(  \Psi(\vec{r}) \nabla \Psi(\vec{r})^*
        - \Psi(\vec{r})^*\nabla \Psi(\vec{r})
  \Bigr),
\end{equation}
one can define, as usually, the transmission and reflection probabilities
\cite{buett85}.
Here $\Psi(\vec{r})^*$ denotes the complex conjugate of the 
scattering wave function (\ref{Psi_gen2}).

The transversal component of the density current 
$j_\perp(x_\perp,x_\parallel,x_3)$ is zero in leads,
because $\phi_n(x_\perp)$ are real functions.
The component $x_3$ of the incident density current 
is also zero, either due to the confinement in the third direction, 
like in Cartesian geometry \cite{roxana09}, 
or due to the symmetry reasons like for the cylindrical geometry \cite{racec09}.
What remains is the longitudinal component of the particle density current
$j_\parallel(x_\perp,x_\parallel,x_3)$,
which provides after the integration over the cross section of the lead
with the corresponding measure, $d\Omega$, the very well-known
relations for the transmission and reflection probabilities.
The probability for an electron incident
from the source, $s=1$, on the channel $n$  to be reflected back into the 
source on the channel $n'$ is 
\begin{equation}
R_{nn'}^{(1)}=\frac{k_{1n'}}{k_{1n}} |S_{1n,1n'}^T|^2,
\end{equation}
and the probability to be transmitted into the drain, $s=2$, on the 
channel $n'$ is 
\begin{equation}
T_{nn'}^{(1)}=\frac{k_{2n'}}{k_{1n}} |S_{1n,2n'}^T|^2.
\end{equation}
The reflection and transmission probabilities for the evanescent (closed)
channels are zero.
The total transmission and reflection coefficients for 
an electron incident from reservoir $s=1$ are defined as
\begin{equation}
T^{(1)}=\sum_{n,n'} T^{(1)}_{nn'}, \quad R^{(1)}=\sum_{n,n'} R^{(1)}_{nn'}.
\end{equation}
More detailed properties of the many-channel tunneling and reflection
probabilities are given in Ref. \cite{buett85}, 
but note that our indexes are interchanged with respect to the definitions 
used there. 

\subsection{Current scattering matrix}
Further, we define the energy dependent \textit{current scattering matrix} as
\begin{equation}
{\tilde{\mathbf{S}}}(E) = 
\mathbf{K}^{1/2}(E) \bm{\varTheta}(E) \mathbf{S}(E) \mathbf{K}^{-1/2}(E),
\label{Stilde}
\end{equation}
so that its elements give directly the reflection and transmission probabilities
\begin{equation}
\begin{split}
|\tilde{S}_{1n',1n}(E)|^2  =  R_{nn'}^{(1)}(E), & \qquad
|\tilde{S}_{2n',2n}(E)|^2  =  R_{nn'}^{(2)}(E),\\
|\tilde{S}_{2n',1n}(E)|^2  =  T_{nn'}^{(1)}(E), & \qquad
|\tilde{S}_{1n',2n}(E)|^2  =  T_{nn'}^{(2)}(E). 
\end{split}
\end{equation}

The diagonal ${\varTheta}$-matrix assures that the matrix elements of
$\tilde{\mathbf{S}}$ are nonzero only for the conducting channels,
for which the transmitted flux is nonzero.
Using the $R$-matrix representation of $\mathbf{S}$, Eq.
(\ref{R-Srelation}), we find from the above relation
\begin{equation}
{\tilde{\mathbf{S}}}(E) = \bm{\varTheta}(E)
                \left[ \mathbf{1} - 2 (\mathbf{1} + i \bm{\varOmega}(E))^{-1}
                \right]
                \bm{\varTheta}(E),
\label{Stilde2}
\end{equation}
with the infinite dimensional matrix $\bm{\varOmega}$
\begin{equation}
\bm{\varOmega}(E) 
=\mathbf{K}^{1/2}(E)\mathbf{R}(E)\mathbf{K}^{1/2}(E)
= u_0 \sum_{l=1}^{\infty}
                      \frac{\vec{\alpha}_l \, \vec{\alpha}^T_l}
                           {E-E_l}
\label{omega}
\end{equation}
and the column vector
\begin{equation}
\vec{\alpha}_l(E) =\mathbf{K}^{1/2}(E) \, \vec{\chi}_l,
\label{alpha}
\end{equation}
with $l \ge 1$. 

Further we express the total tunneling coefficient
in terms of the current transmission matrix,
\begin{equation}
T(E) = \mbox{Tr} [\si(E)
                               \si^\dagger(E)],
\label{T_tr_def}
\end{equation}
where $\si$  denotes the part of $\tilde{\mathbf{S}}$
which contains
the transmission, $\si_{nn'}(E)=\tilde{S}_{2n',1n}(E)$,
$n=\overline{1,N_1(E)}$ and $n'=\overline{1,N_2(E)}$.

According to the definition (\ref{omega}) the matrix
$\bm{\varOmega}$ is a symmetrical one,
$\bm{\varOmega}=\bm{\varOmega}^T$,
and from Eq. (\ref{Stilde2}) it follows that ${\tilde{\mathbf{S}}}$ 
also has this
property, ${\tilde{\mathbf{S}}}={\tilde{\mathbf{S}}}^T$. On this basis one can
demonstrate that the tunneling coefficient characterizes one pair of
open channels 
irrespective of the origin of the incident flux
$T^{(1)}_{nn'}=\left|{\tilde{S}}_{2n',1n}\right|^2
       =\left|{\tilde{S}}_{1n,2n'}\right|^2 = T^{(2)}_{n'n}$.
This is a well-known property of the transmission through a scattering
system and it shows that the current scattering matrix used here is 
properly defined.
The restriction of $\tilde{S}$-matrix to the open channels is the well
known current scattering matrix \cite{wulf98,roxana01,racec02},
commonly used in the
Landauer-B\"uttiker formalism. For a given energy $E$ this is a
$(N_1+N_2) \times (N_1+N_2)$  matrix
which has to satisfy the unitarity
condition, 
according to the flux conservation.

In the numerical computations, 
the matrixes $\mathbf{S}$, $\mathbf{R}$, $\bm{\varOmega}$, 
$\tilde{\mathbf{S}}$ and $\bm{\varTheta}$
have the dimension $ 2N\times 2N$,
and the vectors $\vec{\chi}_l$, $\vec{\alpha}_l(E)$ 
have $2N$ components, where $N$ is the number of scattering channels
(open and closed) taken numerically into account. 
The number of the Wigner-Eisenbud
functions and energies computed numerically 
establishes the maximum value for the index $l$.

\subsection{Resonances}

The relation (\ref{Stilde2}) is the starting point for a resonance theory
of the transmission through a structure with a scattering
region \cite{roxana01,roxana09}.
The singularities of the current scattering matrix $\tilde{\mathbf{S}}$
which satisfy the equation
\begin{equation}
\det \left[ \mathbf{1} + i \bm{\varOmega}(E) \right]=0
\label{res_1}
\end{equation}
are usually classified as bound states and resonances.
The bound states are characterized by real negative energies while
the resonance  energies,
$\bar{E}_{0l}=E_{0l} - i \Gamma_l/2$, $l \ge 1$, lie
in the complex energy plane below the real positive  axis
according to the causality \cite{bohm}.
The scattering matrix $\tilde{\mathbf{S}}$ and consequently the total
transmission $T(E)$ are defined only for
energies in the continuum spectrum (E real positive) of the scattering
problem and they are  analytical functions over the whole domain.
Although they have no singularities in the definition domain,
their energy dependence is determined by the resonances, especially
by those ones which lie in the vicinity of the real axis.
In the resonance domain, i.e. inside a circle of radius $\Gamma_l$
around $\bar{E}_{0l}$, the elements of the current scattering matrix
$\tilde{\mathbf{S}}$ vary strongly with the energy.
In turn, $T(E)$ has also an important variation for the real energies
included in the resonance domain. Thus the resonances appear usually as peaks
in the tunneling coefficient and can be directly seen in the transport
properties of the structure. While in the case of a 1D scattering potential
the peaks are light asymmetric maxima \cite{roxana01},
for a 2D scattering potential the peak shapes cover all ranges 
of the Fano lines,
from asymmetric maxima through "S-type" Fano lines up to 
antiresonances. These profiles have been already seen experimentally
for example in the conductance
of a single-electron transistor \cite{goeres}. In the next section we 
demonstrate 
that the two-dimensional character of the scattering potential and the strong
coupling of the quantum system to the contacts allow for the transmission
profiles which are far from Breit-Wigner lines.

The representation of the $\tilde{S}$-matrix in terms of
$\bm{\varOmega}$, Eq. (\ref{Stilde2})
allows for an efficient numerical procedure
to determine its poles and the resonances.
When the quantum system, for example a quantum dot,
is coupled to the contacts it becomes open, and the
real eigenenergies of the closed problem, $E_l$, migrate in the lower part
of the complex energy plane, becoming resonant energies,
$\bar{E}_{0l} = E_{0l} - i \Gamma_l/2$, $l \ge 1$.
On the base of this correspondence
we fix an energy $E_{\lambda}$ of the isolated dot and determine the
resonance energy $\bar{E}_{0 \lambda}$ as a solution of
Eq. (\ref{res_1}) in the complex energy plane.
The matrix $\bm{\varOmega}$ contains contributions from
all Wigner-Eisenbud functions and energies, i.e. $\chi_l$ and $E_l$, and
from all scattering channels, i.e. all
matrix elements of $\mathbf{K}$.
Thus the resonance energy
$\bar{E}_{0 \lambda}$ can strongly differ from $E_\lambda$,
and only in the case of a very low coupling of the dot to the contacts the
eigenenergies of the isolated system, $E_\lambda$, can properly approximate
the real part of the resonance energy.

The resonance theory presented above
is general and can be applied to a variety of structures
with a 2D scattering potential, regardless if the geometry of the system is
Cartesian or cylindrical. The information about the geometry is contained
only in the Wigner-Eisenbud functions and energies. After solving the
eigenvalue problem of the closed counterpart of the scattering system
one can construct the scattering matrix and analyze it 
without bearing in mind the geometry of the system.

The expression (\ref{Stilde2}) of the $\tilde{S}$-matrix shows
that all matrix elements of $\tilde{\mathbf{S}}$ are singular at the
resonance energy. That means that all transmission coefficients $T_{nn'}$
between different scattering channels
have a similar dependence on energy
around a resonance, and it is enough to analyze the total transmission
which is
a sum of them
in order to characterize the resonance.

\section{Model systems}

Further we analyze the total tunneling coefficient $T(E)$ for a
large energy interval in the case of a quantum dot isolated inside of a
2DEG (\textit{Cartesian geometry}) and in the case of a conical quantum dot
in a cylindrical nanowire (\textit{cylindrical geometry}).
The transmission peaks are directly connected to the resonances
and they have different profiles depending on the coupling strength
between the quantum system and contacts, but also between resonances.

\subsection{Quantum dot in two-dimensional electron gas}

\begin{figure}[h]
\begin{center}
\noindent\includegraphics*[width=4.5in]{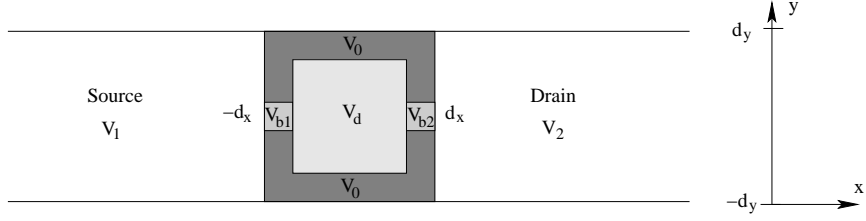}
\end{center}
\caption {
Potential energy in the 2D quantum wire: constant potential energy in the
source and drain contacts, $V_1 \simeq V_2$, and position
dependent potential energy in the dot-region. The quantum dot is isolated
inside the quantum wire by the barrier with the height $V_0$.
The coupling between dot and contacts is set by the potential energy
in the aperture regions, $V_{b1}$ and $V_{b2}$.
The electrons inside the dot experience the potential energy $V_d$.
}
\label{s_pot_2}
\end{figure}

We consider here a quite simple dot, a square dot,
isolated inside a quantum wire by the constant barriers $V_0$ as seen in
Fig. \ref{s_pot_2}.
The smaller barriers $V_{b1}$ and $V_{b2}$ characterize the
coupling between the quantum dot and contacts, and the strength of this
coupling can be varied individually.
Although our model allows for an arbitrary form of the potential, we have
chosen this square dot in order to compare the scattering functions at the
resonant energies with the eigenfunctions of an isolated dot ($V_0$, $V_{b1}$,
$V_{b2}$ $\rightarrow \infty$).

For the numerical calculations we have set $d_x=d_y=50$ nm,
and the width of all barrier $20$ nm. Thus the region where the electrons 
are localized is about $60 \times 60$ nm. The barrier which isolates the 
quantum dot inside the quantum wire has been taken as $V_0=0.4$ eV,
and the potential energy
in the aperture regions $V_{b1}=V_{b2}=0.005$ eV.
In the source and drain contacts the potential energy 
has been considered as the energy reference, $V_1=V_2=0$ eV. 
The Fermi energy of the 
electrons has been taken as  $E_F=12\;$ meV. For this quantum dot there 
are four open channels, $N_1(E_F)=N_2(E_F)=4$, and the closed channels do not
have a significant contribution to the conductance.

The quantum dot described above is strongly coupled to the source and drain 
contacts because the potential energy in the aperture regions lies under 
the Fermi energy. For explaining the transport phenomena 
through the dot it is necessary to take properly into account the
open character of the system and to analyze the transport properties
in terms of the resonances. The eigenstates which characterize 
the closed counterpart of the open dot have an infinite life time
and can not explain the broaden peaks which are experimentally 
measured in the conductance of an open quantum dot \cite{goeres}.

\begin{figure}[h]
\begin{center}
\noindent\includegraphics*[width=100mm]{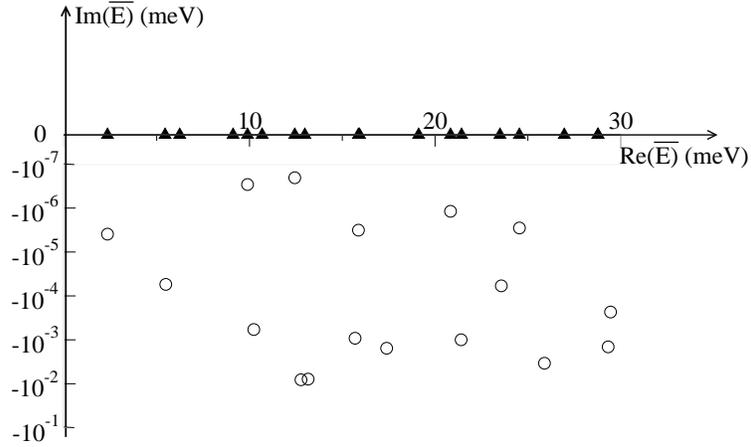}
\end{center}
\caption {
Resonance energies $\bar{E}_{0l}$ (empty circles) of the open quantum dot 
given in Fig. \protect\ref{s_pot_2}
and the real eigenenergies (filled triangles) of the isolated 
counterpart of the considered quantum dot.
}
\label{fig-poles}
\end{figure}

The resonance energies of the considered dot, solutions of Eq. (\ref{res_1}),
are presented in Fig. \ref{fig-poles}.
For comparison, the Wigner-Eisenbud energies, 
i.e. the eigenenergies of the corresponding isolated dot are also given.
Due to the coupling of the quantum dot to the contacts
the resonance energies migrate in the lower part of the complex
energy plane and have different widths.
There are very narrow resonances associated
with the modes of the dot, which are not so strongly coupled to the contacts, 
and broad resonances, which describe modes strongly perturbed by the 
interaction with the reservoirs. 
For a better understanding of the resonance modes
we will examine the localization probability distribution density
of the electrons
for the energies given by the real part
of the resonance energy $\bar{E}_{0 l}$.

\begin{figure}[t]
\includegraphics*[width=4.in]{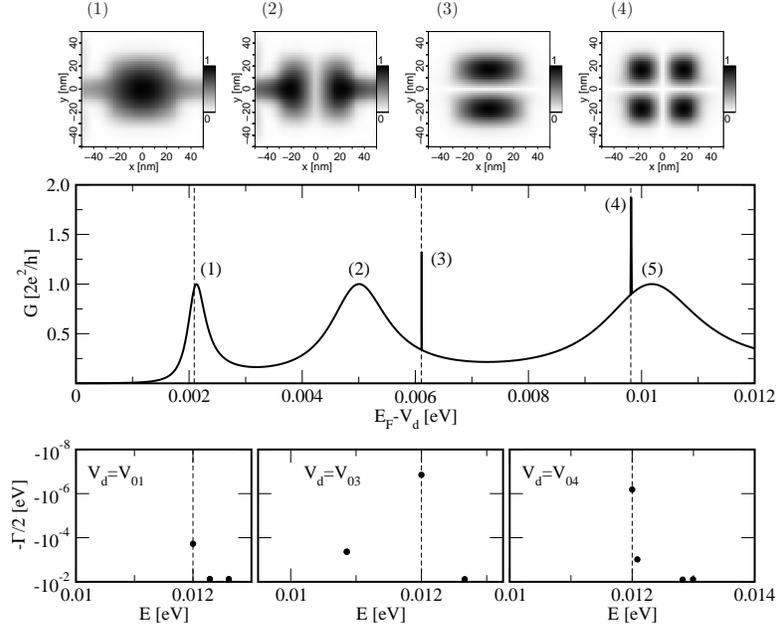}
\caption{(Middle part) Conductance (solid line) as a function of the 
potential energy in the dot region, $V_d$. The vertical dashed lines 
give the position of the potential energy $V_{0l}$, $l=1,3,4$.
(Upper part) The space dependence of the 
electron probability distribution density $P_n(x,y)/P_{max}$, 
$P_n(x,y)=|\psi_n^{(s)}(E;x,y)|^2$, $P_{max}=\max[P_n(x,y)]$, $n=1$ or $n=2$, 
for $E=E_F$ and $V_d=V_{0l}$, $l=1,2,3,4$. 
For the peaks $l=1,2$ we have considered  
the channel number $n=1$, while for the peaks
$l=3,4$, $n=2$. 
(Lower part) Resonance energies in the complex energy plane 
with the real part around $E_F$. The vertical dashed lines correspond 
to the Fermi energy.
}
\label{GG_I}
\end{figure}

\begin{figure}[t]
\begin{center}
\noindent\includegraphics*[width=4.5in]{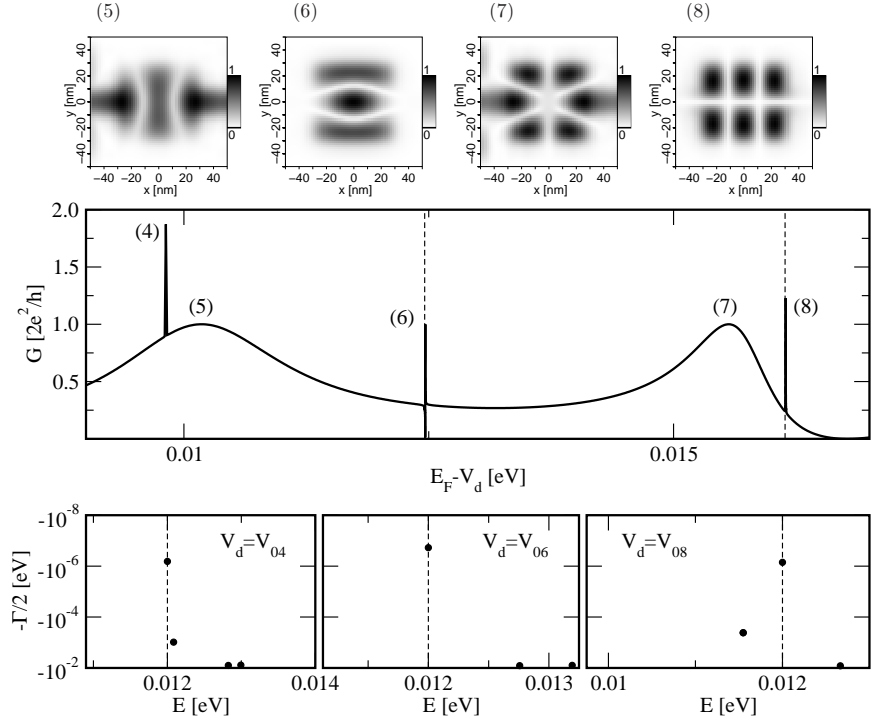}
\caption{(Middle part) Conductance (solid line) as a function of the
potential energy in the dot region, $V_d$. The vertical dashed lines
give the position of the potential energy $V_{0l}$, $l=6,8$.
(Upper part) The space dependence of the
electron probability distribution density $P_n(x,y)/P_{max}$,
$P_n(x,y)=|\psi_n^{(s)}(E;x,y)|^2$, $P_{max}=max[P_n(x,y)]$
for $E=E_F$ and $V_d=V_{0l}$, $l=5,6,7,8$. 
For the peaks $l=5,6,7$ we have considered  
the channel number $n=1$, while for the peaks
$l=8$, $n=2$. 
(Lower part) Resonance energies in the complex energy plane
with the real part around $E_F$. The vertical dashed lines correspond
to the Fermi energy.
}
\label{GG_II}
\end{center}
\end{figure}

The potential energy felt by the electrons inside the dot is $V_d$.
This energy can be modified continuously by varying 
the voltage of a plunger gate \cite{wulf98,roxana01,racec02},
and the conductance through the dot is measured as a function of $V_d$. 
In the linear regime experiments, i.e. small source drain biases,
and for very low temperatures
the conductance $G$ can be directly connected to the total tunneling
coefficient at the Fermi energy \cite{goeres, roxana01,roxana09},
\begin{equation}
G(V_d) = \frac{2 e^2}{h} T(E_F;V_d).
\end{equation}
In the above relation the potential energy $V_d$ appears as a parameter 
in the expression of the total tunneling coefficient.
A new value of $V_d$ means a new scattering potential
and a new scattering matrix. But there is no 
analytic dependence of T on $V_d$, so that  
the scattering matrix and after that the conductance
have been numerically computed for each value of $V_d$.
In Figs. \ref{GG_I} and \ref{GG_II}
the conductance is 
plotted as a function of $E_F-V_d$. 
This fact has the great advantage that
the position of the maxima in conductance are given
with respect
to the bottom of the quantum well, and in this way a direct comparison with
an infinite quantum well is possible.
There are narrow and broad peaks in the conductance
and in order to understand  why they have different profiles
we have also plotted the electron probability distribution density
$P_n(x,y)=|\psi_n^{(1)}(E;x,y)|^2$ for $E=E_F$ and $V_d=V_{0l}$,
for the eight peaks considered here, $l=\overline{1,8}$.
In principle the potential energy $V_{0l}$ is associated with the 
maximum of the conductance peak, 
but the conductance curve shows also a "S-type" 
Fano line, and a rigorous method to fix $V_{0l}$ is necessary.

Using the R-matrix representation of the 
S-matrix, Eq. (\ref{Stilde2}) we can provide an approximative relation 
for $T(E_F,V_d)$ around a resonance
\begin{equation}
T(E_F,V_d) = T(E_F,V_{0l}+\delta V) \simeq T(E_F-\delta V,V_{0l})
\label{T_appr}
\end{equation}
where $V_{0l}$ is the value of $V_d$ for which the real part of
the resonance energy (${\bar{E}}_{0l}=E_{0l}-i \Gamma_l/2$) 
matches the Fermi energy, $E_{0l}=E_F$, 
and $\delta V$ is a small variation 
with values in the interval $(-\Gamma_l,\Gamma_l)$. For a detailed discussion
of this approach see Ref. \cite{roxana01}, Appendix A.
The expression (\ref{T_appr}) of the total tunneling coefficient allows for
a direct connection to the resonances. We can simultaneously plot
the conductance as a function of $E_F-\delta V$, 
$\delta V \in (-\Gamma_l,\Gamma_l)$ and the resonances
with the real energies in the interval $(E_F-\Gamma_l,E_F+\Gamma_l)$.
These plots are given in Figs. \ref{GG_I} and \ref{GG_II}.
The dashed vertical lines correspond in each picture to $\delta V=0$,
i.e. $V_d=V_{0l}$ in the plots in the middle part and 
$E=E_F$ in the plots in the lower part.
From the simultaneous analysis of these graphics 
it is evident that we can associate each peak 
in the conductance with a resonance $l$.
At the resonance energy the electrons show a strong localization 
in the dot region as seen in the upper part of the figures.
A narrow peak corresponds to 
a resonance energy with a very small imaginary part and to a resonance state 
that is almost decoupled from the contacts, i.e the electron probability 
distribution density is nearly zero in the aperture regions. 
These are the resonances 
denoted by (3), (4), (6) and (8) in Figs. \ref{GG_I} and \ref{GG_II}.
The other peaks are broaden  and they correspond to resonances 
with a larger imaginary part and to states which are strongly coupled 
to the contacts. 
Generally, the peaks for which the resonance states have a maximum at $y=0$
couple strongly to the contacts and become broaden. 
These are the peaks (1), (2), (5) and (7) in Figs. \ref{GG_I} and \ref{GG_II}.
But there is an exception which does not depend on the parameters of 
the system: the peak denoted by (6) in Fig. \ref{GG_II}.
Although the probability distribution density has a maximum in the central 
region of the dot, this state is strongly localized inside the dot,
and the corresponding peak is very narrow. But this peak is not an
asymmetric maximum anymore, it has a "S-type" Fano line shape.
This behavior can be explained only taking into account 
the interaction between resonances. The state which corresponds to 
the peak (5) has three maxima in the $x$-direction and one maximum in the
$y$-direction, while the peak (6) has one maximum on $x$- and three
maxima on $y$-direction. They are states with the same symmetry 
in the both directions and they influence each other. 
As a result, there are two hybrid modes, 
one of them very strongly coupled to the contacts
and the other one almost isolated. This interaction
between resonant states with
the same symmetry in the lateral direction is a general
phenomenon which has at the origin the scattering between different 
energy channels due to the nonseparable character of the scattering potential.
For systems with an effective 1D scattering potential the interaction 
between resonances is weak and  the strong
asymmetric Fano line shapes ("S-type" or antiresonance) do not appear.
Also the hybrid modes do not exist in this case.

\begin{figure}[h]
\begin{center}
\noindent\includegraphics*[width=4.5in]{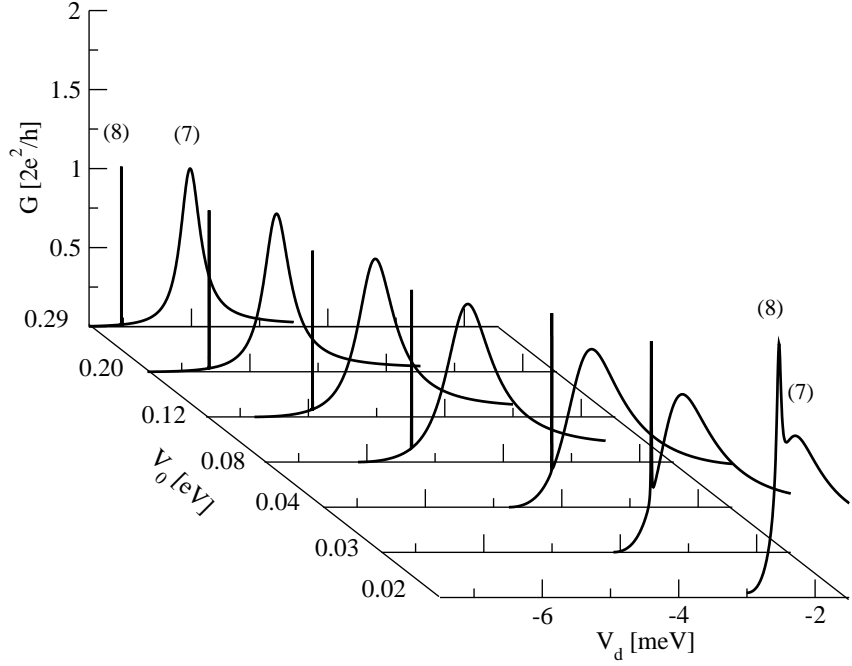}
\caption{
Conductance as a function of the potential energy in the dot region, $V_d$,
for different values of the lateral confinement potential, $V_0$.}
\label{comp}
\end{center}
\end{figure}

Besides the potential energy in the dot region $V_d$, the confinement potential of the dot $V_0$ can be also modified by the top gates. Decreasing the
strength of the confinement the coupling of the resonance states
to the contacts increases, and the conductance peaks become broader.
We present in Fig. \ref{comp} the evolution of the conductance
peaks (7) and (8) when the potential $V_0$ decreases. The width of the
two peaks increases, and the maximum of each peak migrates to higher
energies. But the shift in energy is different. 
The state (7) which is strongly coupled to the contacts in the aperture
regions is much less influenced by the variation of $V_0$. 
The state which corresponds to the peak (8), 
with nearly zero
probability distribution density in the region of the apertures,
can have a coupling to the
contacts only in the case of a rather transparent confinement barrier $V_0$.
This explains the significant broadening of the line shape and 
also the larger shift of the peak energy.

\subsection{Conical quantum dot inside a cylindrical nanowire}

\begin{figure}[b]
\begin{center}
\subfigure[]{\label{1ConDot_sketch}
           \includegraphics*[width=2.75in]{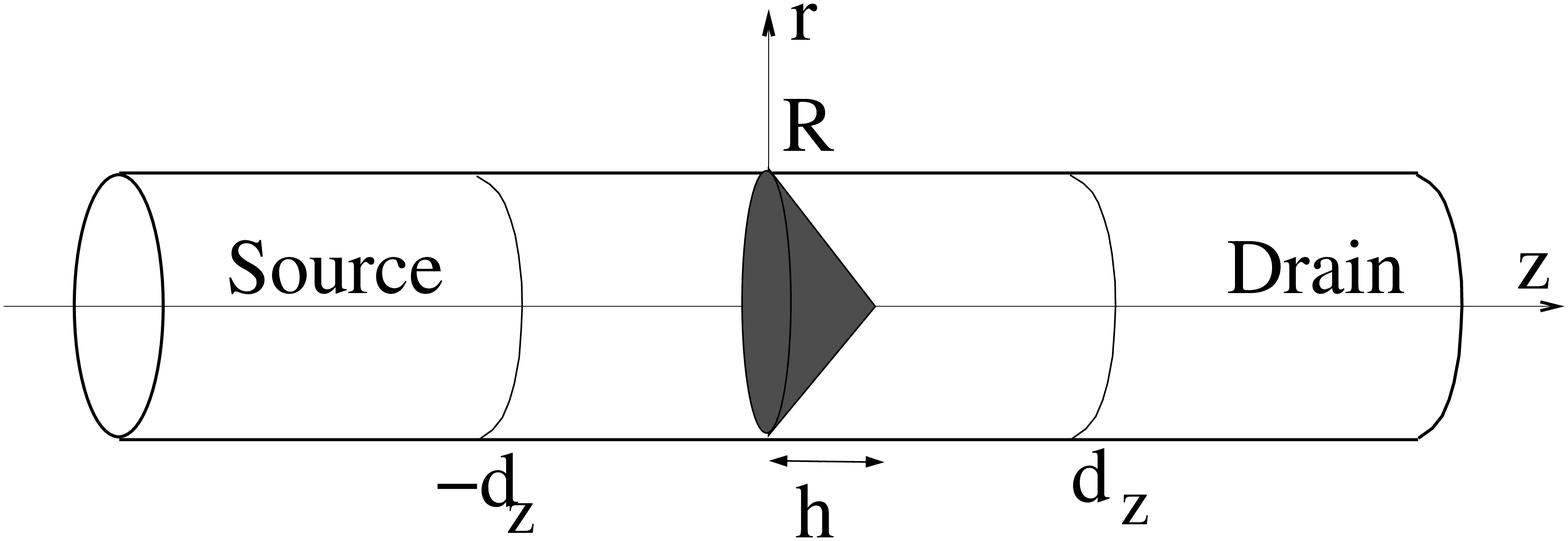}}
\subfigure[]{\label{1ConDot_v_pot}
           \includegraphics*[width=2.5in]{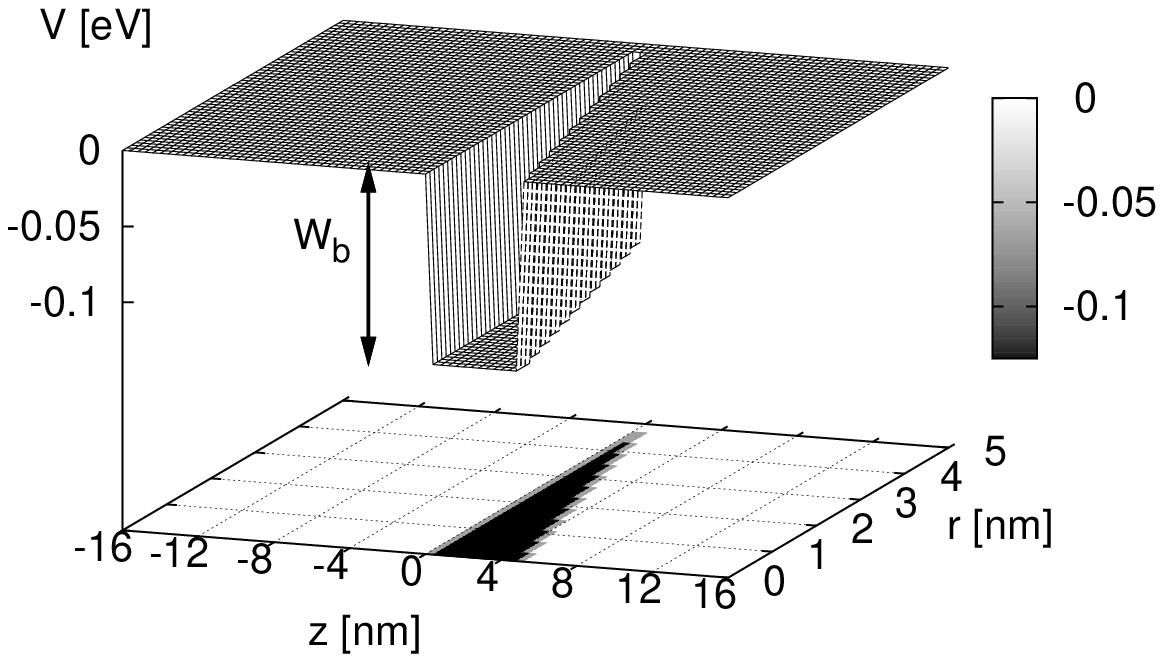}}
\caption{(a) Sketch of a conical quantum dot embedded
into a nanowire with the same radius.
The dot yields an attractive potential $V(z,r)$, represented
in (b) by a quantum well of depth $W_b=-0.125$eV.
}
\label{1ConDot_v}
\end{center}
\end{figure}

We consider a conical quantum dot, 
embedded in an infinite cylindrical nanowire with the same radius,
as is sketched in Fig. \ref{1ConDot_sketch}.
The parameters considered are
the height of the dot $h=5$nm,
the radius of the nanowire $R=5$nm 
and the effective mass $m^*=0.19m_0$. 
We set in our computations $d_\parallel=d_z=16$nm
and the total number of channels (open and closed)
$N=8$. 
In our calculations, the results do not change if more channels are added.

Depending on the 
band-offsets between the dot material and the host material the potential 
produced by the dot can be repulsive, yielding a quantum barrier, or 
attractive, yielding a quantum well. 
We consider here 
that the dot yields an attractive potential $V(z,r)$, represented
in Fig. \ref{1ConDot_v_pot} by a quantum well of depth 
$W_b=-0.125$eV.

\begin{figure}[h]
\begin{center}
\includegraphics*[width=3.in]{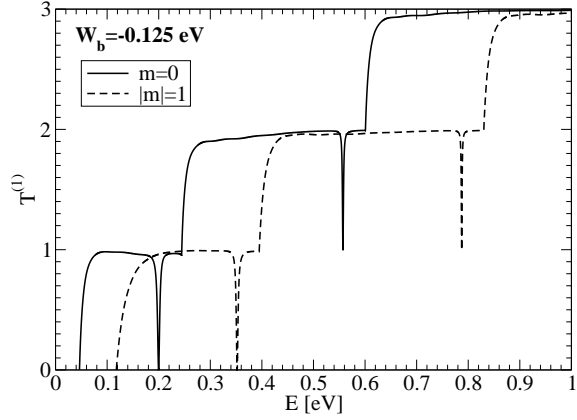}
\caption{Total tunneling coefficient 
as a function of incident energy $E$
for the scattering potential represented in Fig. \protect\ref{1ConDot_v_pot},
for different magnetic quantum numbers 
$m$ (continuous line for $m=0$, dashed line for $|m|=1$) 
and for the well depth $W_b=-0.125$eV. 
}
\label{1ConDot_T1}
\end{center}
\end{figure}

The total tunneling coefficient $T^{(1)}$ versus the incident energy $E$
is plotted in Fig. \ref{1ConDot_T1} for different magnetic
quantum numbers $m$.
The transmission increases with a unity, every time
a new channel $E^{(m)}_{\perp,n}$ becomes available for transport, i.e. 
becomes open. The length of the plateaus is given by the difference between
two successive transversal mode energies, and this length increases with $m$.
Due to the presence of the quantum well, 
deviations appear from the step-like transmission. 
Just before a new channel gets open
there is a dip, i.e. sharp drop,  in the tunneling coefficient.
These dips are owing to modification of the tunneling coefficient due to the
evanescent (closed) channels \cite{bagwell90}. This is a multichannel effect
that was also put in evidence in Cartesian 
coordinates for quantum wires tailored in a two-dimensional electron gas
\cite{bagwell90,levinson93,noeckel94,gudmundsson04,gudmundsson05}. 

The dips can be understood considering 
the simple couple-mode model \cite{bagwell90,levinson93,noeckel94,racec09}.
For a dot surrounded by the host material, 
the scattering potential $V(z,r)$ is not anymore separable,
so that the scattering mixes the channels 
\cite{bagwell90,levinson93,noeckel94,racec09}.
As soon as the scattering potential is attractive, 
the diagonal coupling matrix element 
\begin{equation}
V_{nn}(z)=\int_0^R \phi_n(r)V(z,r)\phi_n(r) r dr < 0
\label{eff_attr}
\end{equation}
acts for every channel $n$ as an 
effective one-dimensional (1D) attractive potential \cite{levinson93}, 
which always allows for at least one bound state 
\cite{simon76,klaus77}
below the threshold of the continuum spectrum. 
By mixing the channels, this bound state becomes a 
\textit{quasi-bound state} or resonance, 
i.e with complex energy, whose real part gets embedded into continuum 
spectrum of the lower channel and the imaginary part describes
the width of the resonance. 
These resonances can be seen now as dips in the tunneling coefficient.
The energy difference between the position of the dips 
and the next subband minima $E^{(m)}_{\perp,n}$ 
gives the quasi-bound state energy. 
The positions of the dips, i.e. the quasi-bound state energy, 
depend on the channel number
$n$ and on the magnetic quantum number $m$ and, of course, on the detailed
system parameters.
In Cartesian coordinates the specific symmetry of the channels (odd and even) 
do not allow for dips in the first plateau \cite{gudmundsson04}. 
In the cylindrical geometry
this symmetry is broken, so that we obtain a dip in front of every
plateau.
Our numerical method allows for a high energy resolution in computing the
tunneling coefficient, so that 
we were able to find
the dips also in front of the higher-order plateaus.

\begin{figure}[thb]
\subfigure[$E=0.199$eV, $n=1$]{\label{1ConDot_psi_1}
          \includegraphics*[width=2.in]{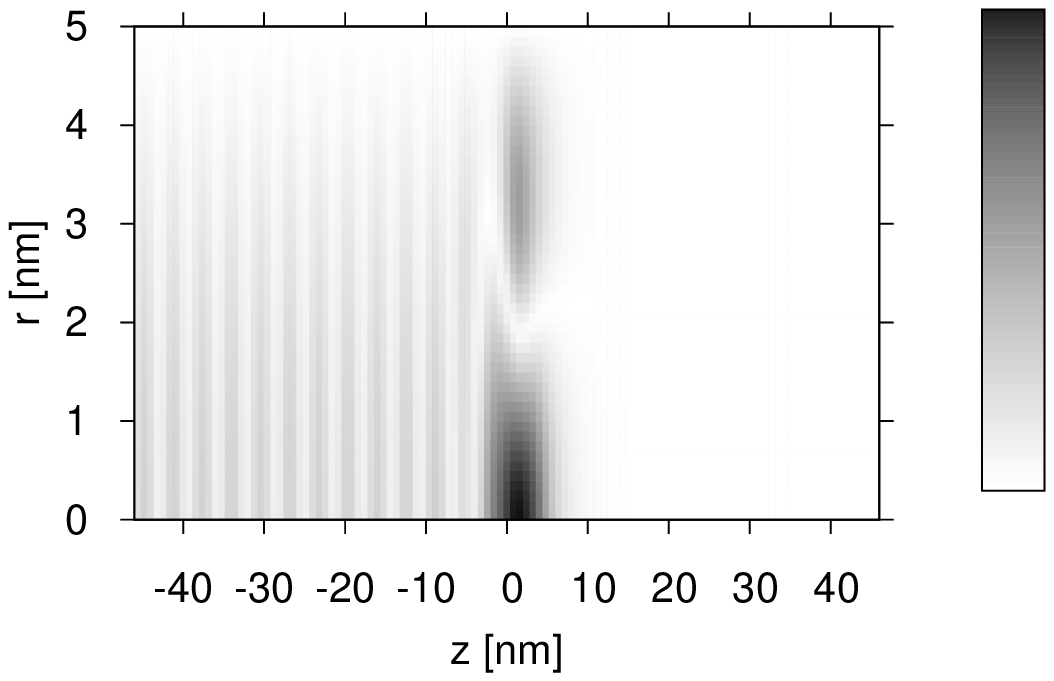}}
\subfigure[$E=0.557$eV, $n=2$]{\label{1ConDot_psi_2}
          \includegraphics*[width=2.in]{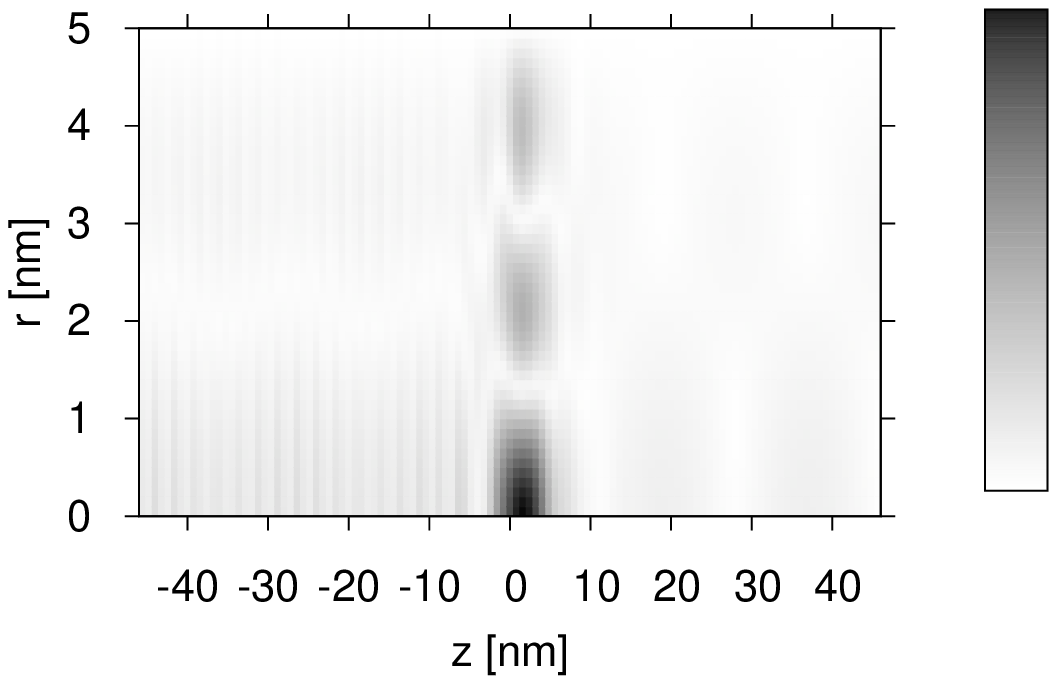}}
\caption{Localization probability distribution density, 
$|\psi^{(1)}_n(E;z,r)|^2$,
for an electron with $m=0$,
incident from the reservoir $s=1$ into the channel $n$ and 
with the total energy $E$.
The energies are the dips in Fig. \protect\ref{1ConDot_T1}. 
}
\label{1ConDot_psi}
\end{figure}

Further insight about the quasi-bound states of the evanescent
channels can be gained looking at the wave functions,
whose square absolute value $|\psi^{(s)}_n(E;z,r)|^2$ gives the 
\textit{localization probability density}.
The R-matrix formalism
allows us to produce high resolution maps of the
wave functions 
inside the scattering region, see Eq. (\ref{psi3}).
In Figs. \ref{1ConDot_psi_1}, 
\ref{1ConDot_psi_2}
the localization 
probability distribution density 
is represented 
in arbitrary units,
for an electron  
incident from source ($s=1$) and with a total energy corresponding
to the dips in Fig. \ref{1ConDot_T1}. 
The total energy $E$ 
and the channel $n$, on which the electron is incident, are specified at every
plot. 
Let discuss Fig. \ref{1ConDot_psi_1}. 
The total energy $E=0.199eV$ is less than
the energy of the second transversal mode, $E_{\perp,2}^{(0)}=0.244eV$,
so that only the first channel is open.
Thus the incident wave from the source contact 
is node-less in $r$-direction. 
But, as it can be seen in Fig. \ref{1ConDot_psi_1}, 
the scattering wave function inside the scattering region 
has a node in the $r$-direction, i.e. position in $r$ where the
wave function is zero. 
This means that the wave function 
corresponds to the quasi-bound state splitting off 
from the second transversal mode,
which is an evanescent one.
The quasi-bound state is
reachable now in a scattering formulation due to channel mixing.  
The scattering wave function has a pronounced peak around the 
scattering potential,
i.e. $z\in [0,5]$nm,
and decreases exponentially to the left and to the right.
On the left side of the scattering potential one observes the interference
pattern produced by the incident and the reflected waves, while
on the right side there exists only the transmitted wave. 

The scattering wave function considered in Fig. \ref{1ConDot_psi_2} 
has the energy less than the third transversal channel,
$E^{(0)}_{\perp,3}=0.6006eV$,
so that the incident part of the scattering state 
on the second mode $n=2$ 
has one node in $r$-direction. But the scattering function shows 
inside the scattering region two nodes in
the $r$-direction, so it corresponds
to a quasi-bound state splitting off from the above evanescent channel,
the third one.

One gets similar pictures for all $m$-values, with the
difference that for $m\not=0$ the wave functions
are zero for $r=0$. 
In Fig. \ref{1ConDot_psi_1} and \ref{1ConDot_psi_2}
one can observe that the transmitted part of the scattering wave function 
is zero, in agreement with the resonant backscattering specific
to the quasi-bound states of the evanescent channels 
\cite{bagwell90,levinson93}.
Increasing the strength of the attractive potential
one can see more dips \cite{gudmundsson04,racec09}
in the tunneling coefficient. 
Another systems 
embedded inside the cylindrical nanowire, 
like a cylindrical dot,
a quantum ring or a double barrier heterostructure, 
which also show a similar behavior, 
were studied in Ref. \cite{racec09}.

\begin{figure}[h]
\subfigure[]{\label{1ConDot_be}
\includegraphics*[width=2.in]{fig10a} }
\subfigure[]{\label{1ConDot_bs}
\includegraphics*[width=2.5in]{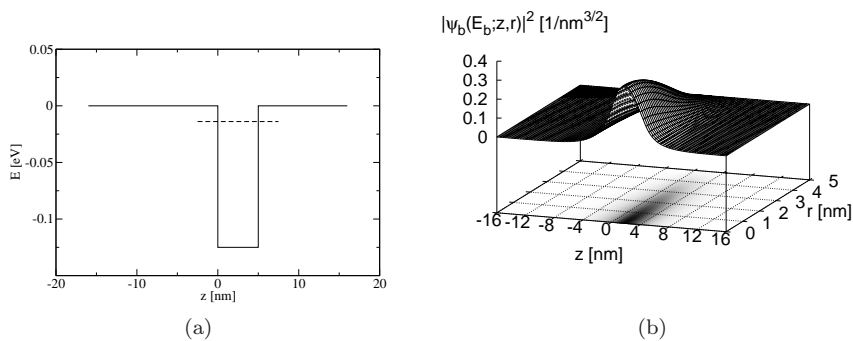} }
\caption{The bound state for a conical quantum dot
inside the nanowire, as in Fig. \protect\ref{1ConDot_v}.
(a) Value of the bound energy $E_b=-0.013$eV (dashed line). 
The potential energy along $r=0$ is represented by the 
solid line.
(b) The absolute value square of the wave function
corresponding to the bound state, $|\psi_b(E_b;z,r)|^2$.
}
\end{figure}

We have analyzed until now only the classical allowed energy domain,
with a continuous, double degenerated spectrum. The wave functions 
of the electrons for these energies are extended states, presented
here as scattering states.
The classically forbidden spectrum contains the bound states
or the localized states. The R-matrix formalism presented here
can provide also these states, as long as the boundary points
$\pm d_z$ are far enough from the quantum dot, so that the bound states 
fulfill the Neumann boundary condition (\ref{neumann_bc}). 
In such a way, the energies of the bound states are the 
negative Wigner-Eisenbud energies  and 
the wave functions for the bound state
are the corresponding Wigner-Eisenbud functions.
For the conical dot presented here,
there is only a bound state represented by the lowest 
Wigner-Eisenbud energy, $E_b=E_1 < 0$ and $\psi_b(E_b;z,r)=\chi_1(z,r)$.
In Fig. \ref{1ConDot_be} we present the energy value $E_b$ of the bound state 
together with a cut of the potential energy along $r=0$.
The corresponding wave function, namely the absolute value square,
is represented in Fig. \ref{1ConDot_bs}.
One can see that the bound state is mainly 
localized inside the quantum dot. 

\section{Summary and discussion}

We have presented a general theory for computing the
scattering matrix and the scattering wave functions
for a general finite-range extended scattering potential
in two dimensions. 
The theory is based on the R-matrix formalism, 
which allows a semi-analytical treatment of the scattering problem,
yielding in such a way a powerful and efficient numerical method. 

This formalism was applied to a quantum dot defined inside a two-dimensional
electron gas, as well to a conical quantum dot embedded inside
a cylindrical nanowire.

It is pointed out the role of the evanescent channels,
which for a nonseparable attractive
scattering potential in a multi-channel nanowire 
produces resonant dips in the tunneling coefficient.
Furthermore, the cylindrical symmetry does not yield the same "selection rules"
for tunneling coefficient as the Cartesian symmetry.

It is also presented a general resonance theory, which shows that
the two-dimensional character of the scattering potential and the strong
coupling of the quantum system to the contacts allow for the transmission
profiles which ranges from asymmetric Fano line shapes, 
through "S-type" Fano lines until antiresonances.

Detailed maps of localization probability distribution density sustain
the physical interpretation of the resonances (dips and peaks) 
found in the studied heterostructures.  

\begin{acknowledgments}
It is a pleasure for us to acknowledge the fruitful discussions with 
Klaus G\"artner, Vidar Gudmundsson, Andrei Manolescu and Gheorghe Nenciu.
One of us (P.N.R.) also acknowledge partial support 
from German Research Foundation through SFB 787 
and from the Romanian Ministry of Education and Research 
through the Program PNCDI2, Contract number 515/2008.
\end{acknowledgments}

\bibliography{nwt}

\end{document}